\documentstyle[aps,manuscript]{revtex}   

\begin{document}

\draft

\title{Fluctuations and the existence of potential in dissipative 
semiclassical systems}

\author{Bidhan Chandra Bag and 
Deb Shankar Ray{\footnote{pcdsr@mahendra.iacs.res.in}} 
}

\address{Indian Association for the Cultivation of Science,
Jadavpur, Calcutta 700 032, INDIA.}

\maketitle

\begin{abstract}
We examine the weak noise limit of an overdamped dissipative
system within a semiclassical description
and show how quantization influences the growth
and decay of fluctuations of the thermally equilibrated systems. 
We trace its origin
in a semiclassical counterpart of the generalized potential for the dissipative 
system.
\end{abstract} 

\vspace{1cm}

\newpage

\section{Introduction}

One of the important issues in nonequilibrium phenomena in the macroscopic
nonlinear systems is to understand the 
interplay of nonlinearity of the system and the fluctuations of its environment. 
The problem is fairly general in the 
context of chemical reactions \cite{dykman2},
breakdown of electronic devices \cite{kautz}, phase transitions \cite{nature}
etc. The essential
description of the physical situation rests on the Fokker-Planck equations for the
probability distribution functions of the relevant variables of the dynamics.
In the weak noise limit the fluctuations have been described 
\cite{nature,bray,graham,tel} by appropriate auxiliary
Hamiltonian or path integral methods. 
The theoretical results have been 
corroborated  by remarkable experiments on fluctuations using analogue 
electronic circuits \cite{nature} , which allow the phase space 
trajectories of fluctuations
to be observed directly in a precise manner. These studies have enriched our
understanding in several theoretical issues, e. g. , the symmetry between 
the growth and the decay of classical fluctuations in equilibrium 
and its breakdown under 
nonequilibrium conditions \cite{nature} , the existence of a nonequilibrium
potential of a dissipative 
system \cite{graham,tel} etc.

It is the purpose of this paper to extend the theory to the semiclassical
context. The quantization of the system itself adds a new dimension to the
interplay of nonlinearity and stochasticity in a dissipative 
system. To make a fair comparison with classical theory
we adopt the Wigner's phase space distribution function of c-number variables.
The weak noise limit can then be appropriately employed to develop an auxiliary 
Hamiltonian formulation at the semiclassical level 
in terms of these phase space variables. This allows us to realize 
the existence of an optimal force of purely quantum origin derivable
from the fluctuating field and relate it to the momentum of the auxiliary
Hamiltonian. The quantum correction also makes its presence felt in the growth
and decay of fluctuations of thermally equilibrated semiclassical 
systems keeping the symmetry preserved.

The outline of the paper is as follows: In Sec. II we introduce the general
aspects of dynamics of dissipative quantum system 
in terms of the Wigner phase space function. 
In Sec. III we consider the weak noise 
and semiclassical limit under overdamped condition and take resort to the well-known auxiliary 
Hamiltonian description. The quantum part of the optimal force is then 
explicitly derived. The symmetry between the
growth and decay of fluctuations in a thermalized quantum system is 
discussed in Sec. IV. The existence of a semiclassical contribution to the potential
in a dissipative system is then shown in Sec. V. The paper is concluded in
Sec. VI.

\section{Quantum dynamics of a dissipative system}

We consider a dynamical system characterized by a potential $V(x)$ coupled to an 
environment. Evolution of such an open quantum system
has been studied over the last several decades under a variety of reasonable 
assumptions \cite{louisell,legg,gg1}. Specifically interesting is the semiclassical limit of an Ohmic 
environment. The dissipative time evolution of the Wigner distribution
function $W(x, p, t)$ for the system with unit mass ($m = 1$) 
under the potential $V(x)$ can be described by \cite{zurek,bag1,pat}

\begin{equation}
\frac{\partial W}{\partial t} = 
-p \frac{\partial W}{\partial x}
+ \frac{\partial V}{\partial x} \frac{\partial W}{\partial p} 
+ \epsilon \sum_{n\geq 1} 
\frac{ \hbar^{2n}(-1)^n}{ 2^{2n}(2n+1)! }
\frac{\partial^{2n+ 1}V}{\partial x^{2n+ 1} }
\frac{ \partial^{2n+1} W }{\partial p^{2n+1} } +
\gamma \frac{\partial p W}{\partial p}  
+ D \frac{\partial^2 W}{\partial p^2}  \; \;,
\end{equation} 

\noindent
where $\gamma$ and $D$ are the dissipation constant and the diffusion 
coefficient, respectively. $x$ and $p$ are c-number co-ordinate and 
momentum variables. The drift term is a direct consequence \cite{legg} of the 
existence of $\gamma$-dependent term in the imaginary part of the exponent
in the expression for the propagator for the density operator in the
Feynman-Veron theory and has been shown to be responsible for appearance of a
damping force in the classical equation of motion for the Brownian particle
to ensure quantum-classical correspondence. $\gamma$ and D are related by the
fluctuation-dissipation relation, $D = \frac{\gamma}{2} \hbar \omega
\coth{\frac{\hbar \omega}{2 k_b T}}$ (in the semiclassical 
limit  $D = \gamma k T$ ). $\omega$ is the renormalised linear frequency
of the nonlinear system.
The quantum correction to classical Liouville motion is contained in the
$\hbar$-containing terms in the sum. $\epsilon$ is a parameter (whose value is 1) which is 
kept in the equation for bookkeeping the Wigner correction term in our further
analysis. We put $\epsilon=1$ at the end of calculation. 

Eq.(1) had been used earlier in several occasions. For example, Zurek and Paz
\cite{zurek} and others \cite{pat} have studied some interesting aspects of quantum-classical
correspondence in relation to decoherence and chaos. Based on this equation 
and its variant chaotic dissipative systems has been studied.
\cite{bag1,mil,dit,bag2}. The equation also yields the simplest leading order
quantum correction term to classical Kramers' rate \cite{jr}. 
The primary reason for choosing Eq.(3) as our starting point is 
that it reaches the correct classical limit when $\hbar \rightarrow 0$ so that
$D$ becomes a thermal diffusion coefficient ($\gamma k T$) in the high 
temperature limit and the Wigner function reduces to the corresponding classical
phase space distribution function
and we recover the Kramers' equation which describes classical Brownian
motion of a particle in phase space.

\section{Weak noise and semiclassical limit of quantum dissipative dynamics:}

The weak noise limit of a dissipative system within 
a semiclassical description can
be conveniently described by a ``WKB-like'' ansatz (we refer to ``WKB-like'' since we are 
considering more than one dimension. Traditionally WKB refer to one dimension
only)
of the Eq.(1) for the Wigner function of the form

\begin{equation}
W(x, p, t) = Z(x, t) \exp(-\frac{x p}{\hbar}) \exp(-\frac{s}{D_1}) \; \;.
\end{equation}

where $D_1 = \frac{D} {\omega^2}$. The weak noise limit is defined \cite{nature}
as $D_1 \rightarrow 0$ and semiclassical limit refers to ($\hbar \rightarrow 0$). 
$Z(x,t)$ is a prefactor and $s(x, p, t)$ is a classical action which is a
function of c-number variables $x$ and $p$ , satisfying the following 
Hamilton-Jacobi equation,

\begin{eqnarray}
& & \frac{\partial s}{\partial t} + p \frac{\partial s}{\partial x} - 
V' \frac{\partial s}{\partial p} - \gamma p \frac{\partial s}
{\partial p} 
+ \omega^2 (\frac
{\partial s}{\partial p})^2 \nonumber\\
& & + \epsilon
\sum_{n\geq 1} \frac{x^{2n} (-1)^{3n+1}}{2^{2n}(2n)!} \frac{\partial^{2n+1} V}
{\partial x^{2n+1}} \frac{\partial s}{\partial p} = 0 \; \;.
\end{eqnarray}

The derivation of Eq.(3) is based on the following consideration. Since in the
weak noise limit $D_1$ is the relevant small parameter one obtains with 
ansatz (2) in leading order a term proportional to $\hbar^{2n} \left( \frac{1}
{D_1} \frac{\partial s}{\partial p} \right)^{2n+1}$ which is not balanced by
any other term of the same order ${D_1}^{-(2n+1)}$. This is because the 
highest derivative in Eq.(1) does not have a factor scaling with the 
corresponding power of $D_1$. The successive terms next to the leading order are
also $\hbar$-containing terms. All of these terms vanishes in the semiclassical
limit ($\hbar \rightarrow 0$). Therefore the leading order term that remains gives
rise to Eq.(3). It is thus obvious that an ansatz (2) with $\hbar$ finite
is not feasible. The semiclassical limit  $\hbar \rightarrow 0$ is a necessary 
requirement for the validity of ansatz (2).
The above equation can be solved by integrating the Hamiltonian equations of
motion,

\begin{eqnarray}
\dot{x} & = & p \nonumber\\ 
\dot{X} & = & P -  \gamma X \nonumber\\
\dot{p} & = & V' + \gamma p - 2  \omega^2 X - \epsilon 
\sum_{n\geq 1}  \frac{ (-1)^{3n+1} x^{2n} }{2^{2n}(2n)! } 
\frac{\partial^{2n+1} V}{\partial x^{2n+1}}  \nonumber\\
\dot{P} & = & \left[V'' - \epsilon 
\sum_{n\geq 1} \frac{(-1)^{3n+1}}{2^{2n}(2n)!} 
\frac{\partial}{\partial x} \left( x^{2n} 
\frac{\partial^{2n+1} V}{\partial x^{2n+1} } \right) \right] X
\end{eqnarray}

which are derived from the following effective Hamiltonian $H_{eff}$

\begin{eqnarray}
H_{eff} & = & p P - V' X 
- \gamma X p 
+ \omega^2 X^2  \nonumber\\
& & + \epsilon
\sum_{n\geq 1} \frac{x^{2n} (-1)^{3n+1}}{2^{2n}(2n)!} 
\frac{\partial^{2n+1} V}{\partial x^{2n+1}}  X \; \; .
\end{eqnarray} 

\noindent
Here we have put $\frac{\partial s}{\partial x} = P$ and
$\frac{\partial s}{\partial p} = X$.
The introduction of additional degree-of-freedom by incorporating the 
auxiliary momentum (P) and co-ordinate (X) makes the system an effectively 
two-degree-of-freedom system. The origin of these two variables is 
the thermal fluctuations of the environment \cite{nature}. 
The auxiliary Hamiltonian $H_{eff}$ is not to be 
confused with the microscopic Hamiltonian comprising the system, the bath and 
their coupling. Thus the phase space 
trajectories concern fluctuations of the c-number variables.

Under overdamped condition ($\ddot{x} \ll \gamma \dot{x} ; \ddot{X} 
\ll \gamma \dot{X}$) Eqs. (4) can be easily reduced to the following form ;

\begin{eqnarray}
\dot{x} & = & K(x) + \frac{2 \omega^2 X}{\gamma} \nonumber\\ 
\dot{X} & = & - \frac{\partial K(x)}{\partial x} X 
\end{eqnarray}

\noindent
where
\begin{equation}
K(x) = \frac{1}{ \gamma} \left[-V' + \epsilon 
\sum_{n\ge 1} \frac{(-1)^{3n+1} x^{2n}}{2^{2n}(2n)!} 
\frac{\partial^{2n+1} V}{\partial x^{2n+1} }  \right] 
\end{equation}

It is easy to recognize the quantity $\frac{2 \omega^2 X}{\gamma}$ as a 
momentum and redefine it as $p_r$. Therefore Eqs(6) may
be rewritten as

\begin{eqnarray}
\dot{x} & = & K(x) + p_r  \nonumber\\
\dot{p_r} & = & \frac{\partial K(x)}{\partial x} p_r  \; \; .
\end{eqnarray}

So the overdamped motion is described by the following effective Hamiltonian,

\begin{equation}
H_{od} = \frac{p^2_r}{2} + K(x) p_r  \; \; .
\end{equation}

The above auxiliary Hamiltonian description (8, 9) is isomorphic in form
to that of Luchinsky and McClintock \cite{nature}, who had considered 
an overdamped classical Brownian motion in a force field $K(x)$, driven by a weak 
white noise $\zeta(t)$ whose intensity $D_1 \ll 1$ as

\begin{equation}
\dot{x} = K(x) + \zeta(t), \; \; \; \langle \zeta \rangle = 0 \; \; ,
\; \langle \zeta(t) \zeta(0) \rangle = D_1 \delta(t) \; \; \; .
\end{equation}

Equivalently the corresponding Fokker-Planck equation for the probability
density $P(x, t)$ is

\begin{equation}
\frac{\partial P(x, t)}{\partial t} = - \frac{\partial K(x)}{\partial x} P(x, t)
+ \frac{D_1}{2} \frac{\partial^2 P(x, t)}{\partial x^2} \; \;.
\end{equation}

The large fluctuations of scale $\gg \sqrt {D_1}$ can therefore be treated 
in the weak noise limit $D_1 \rightarrow 0$ \cite{nature,graham,tel}
by ``WKB  like'' approximation of the Fokker-Planck equation (11) in the 
form 

\begin{equation}
P(x, t) = y(x, t) \exp[ \frac{- \phi(x,t)}{D_1}] \; \;.
\end{equation}

Here y(x, t) is the prefactor and $\phi(x, t)$ is a ``classical'' action 
describing a Hamiltonian-Jacobi equation which can be solved by solving 
Hamilton' s equation (8) with $p_r = \frac{\partial \phi}{\partial x}$
as the momentum for auxiliary system.

The important distinguishing feature of the above description in which 
the system is treated semiclassically is the structure of $K(x)$ which 
is given by equation(7) and comprises of two terms ;

\begin{equation}
K(x) = K_{cl} + K_{semi}(x)
\end{equation}

\noindent
where

\begin{equation}
K_{cl} = - \frac{V'(x)}{\gamma}
\end{equation}

is derivable from purely classical potential $V(x)$ and $K_{semi}(x)$ 
does not 
explicitly involve $\hbar$,

\begin{equation}
K_{semi}(x) = \frac{\epsilon}{\gamma} 
\sum_{n\geq 1} \frac{x^{2n} (-1)^{3n+1}}{2^{2n}(2n)!} 
\frac{\partial^{2n+1} V}{\partial x^{2n+1}}  
\end{equation}

originates from the nonlinearity of the potential $V(x)$ 
and quantum nature of the system. The quantum
contribution to $K(x)$ is therefore 
likely  to influence both the fluctuational
and the relaxational paths of the dynamics and is also responsible for the
existence of a potential.
Our objective is now to explore these aspects in the following two 
sections.

\section{Large fluctuations in equilibrated semiclassical systems }

In the thermally equilibrated systems a typical large fluctuation of the 
variable $x$ implies a temporary departure from its stable state, $x_s$ to 
some remote  state $x_f$. This is followed by a return to $x_s$ as a result 
of relaxation in the absence of fluctuations $p_r$. A nonzero value of $p_r$  
which results from fluctuations due  
to surrounding drives the system away from $x_s$ along a set of trajectories
which form the unstable invariant manifold and define the so called 
fluctuational paths. On the other hand the system relaxes along the 
relaxational return path to $x_s$ under the condition $p_r = 0$, which form 
stable invariant manifold. The latter condition implies $\dot{x} = K(x)$.
In each case the trajectory represents the optimal paths along which the system 
is expected to move with overwhelming probability. Luchinsky and McClintock \cite{nature}
have studied these paths in analog electronic circuits and demonstrated the 
growth and the decay of {\it classical} fluctuations in equilibrium. We extend
this analysis to semiclassical domain using the same model potential,

\begin{equation}
V(x) = \frac{1}{4} x^4 - \frac{1}{2} x^2 \; \;.
\end{equation}

The quantum contribution to the growth and the decay of fluctuations can be understood
by recognizing the $K_{semi}(x)$ term in the dynamics (8). In Fig.1 we 
compare both the deterministic fluctuational and relaxational (optimal)
paths for quantum and classical thermally equilibrated systems. It is
important to note that the maximum possible amplitude of large fluctuations
is almost double for the quantum system compared to that for the 
corresponding the classical system. This is due to the addition of the 
nonlinear force term of quantum origin, $K_{semi}(x)$ in the 
c-number equation (8), which is shown to be derivable from the 
fluctuating field, and is related to the momentum of the auxiliary Hamiltonian. 

Before leaving this section we make a brief remark on the thermally 
nonequilibrated systems. Since the detailed balance is not operative here,
the optimal path to a given state not just the time-reversed dynamical path 
along which the system moves from this state to the stable state in absence
of fluctuations $p_r$. Thus for the driven system, for example, the pattern of 
optimal path is generically different from that for the thermally equilibrated 
systems. It may display singularities whose topological manifestations as 
caustics, switching line, cusps etc have been thoroughly studied 
for classical systems. We believe that $K_{semi}(x, t)$ where t 
signifies the driving by a periodic force in $V(x, t)$ is likely to play an 
important role in their quantum counterparts.

\section{Existence of a  potential for dissipative semiclassical system}

In a significant analysis Graham and coworkers \cite{graham,tel} had  examined the condition for 
existence of potential for classical dissipative systems. We now extend this analysis
to the present semiclassical context.

The general criterion for a dissipative dynamical system described by 
autonomous equations of the form

\begin{equation}
\dot{x}^{\nu} = K^\nu (x)
\end{equation}

to have a potential $\phi(x)$ with respect to $Q^{\nu \mu}$ (positive, 
semidefinite symmetric matrix, which are considered to be the matrix of
transport coefficients) if there exists a single-valued continuously
differentiable and globally defined function $\phi(x)$, bounded from 
below which is stationary in the limit sets of Eq.(17) and which satisfies

\begin{equation}
K^{\nu}(x) = -\frac{1}{2} Q^{\nu \mu} 
\frac{\partial \phi(x)}{\partial x^\mu}
+ r^{\nu} (x)
\end{equation}

\noindent
with

\begin{equation}
r^{\nu} \frac{\partial \phi (x)}{\partial x^{\nu}} = 0 \; \;.
\end{equation} 

Here for simplicity $Q^{\nu \mu}$ is assumed to be independent of $x$. The 
first and the second terms of Eq.(18) correspond to irreversible and reversible
part, respectively.

The stochastic process $x (D_1, t)$ which involve Eq.(17) and a symmetric 
non-negative matrix $Q^{\nu \mu}$ is governed by the Fokker-Planck equation
for probability distribution function $P(x, t)$

\begin{equation}
\frac{\partial P(x,t)}{\partial t} = - \frac{\partial}{\partial x^{\nu}} 
K^{\nu}(x) P + \frac{D_1}{2} 
\frac{\partial^2}{\partial x^{\nu} \partial x^{\mu}}
Q^{\nu \mu} p \; \;.
\end{equation}

For $D_1 = 0$ the above description reduces to (17). For $D_1 \ne 0$ the 
steady state distribution defines the function $\phi(x, t)$ by 

\begin{equation}
P(x, D_1, t \rightarrow \infty) = N(D_1) \exp [ -\frac{\phi(x, D_1)}{D_1}]
\end{equation}

N is the normalization constant.
If $\phi(x) = \lim_{D_1 \rightarrow 0}  
\phi(x, D_1)$ is a single-valued, continuously differentiable
function bounded from below it satisfies

\begin{equation}
K^{\nu}(x) \frac{\partial \phi(x)}{\partial x^{\nu}}
+ \frac{1}{2} Q^{\nu \mu} 
\frac{\partial \phi(x)}{\partial x^{\nu}}  
\frac{\partial \phi(x)}{\partial x^\mu} = 0 \; \;
\end{equation}

Eq. (22) is equivalent to Eqs. (18, 19). Interpreting Eq.(22) as usual as a
Hamilton-Jacobi equation by defining $\phi(x)$ as action and $\frac{\partial
\phi}{\partial x^{\nu}} = P_{\nu}$ 
as the momentum conjugate to $x_{\nu}$, one 
can construct the auxiliary Hamiltonian

\begin{equation}
H(x,p) = \frac{1}{2} Q^{\nu \mu} p_{\nu} p_{\mu}  + K^{\nu}(x) p_{\nu} \; \;.
\end{equation}

Graham and co-workers \cite{graham,tel} have argued that a potential can exist with equation (17) if the 
above Hamiltonian is integrable for $H = 0$, because the condition implies 
that there exist a smooth separatrix which connect smoothly the stable and unstable
manifolds emanating from the hyperbolic fixed points of the dynamical system.

We now turn back to our dissipative semiclassical system described by Eqs(8)
and (9) where $K(x)$ is defined by Eq.(7). Recognizing Eq.(9) as Eq.(23) 
for a one-degree-of-freedom system we identify 

\begin{eqnarray}
Q & = & 1 \nonumber\\
K(x) & = & K_{cl} + K_{semi} \; \;.
\end{eqnarray}

The potential function $\phi(x)$ can therefore be calculated from Eq.(9) 
with $H = 0$ as (since $p_r = \frac{\partial \phi(x)}{\partial x}$)

\begin{equation}
\phi(x) = -2 \int K(x) dx \; \;.
\end{equation}

The above expression can be made more explicit if we make use of Eq.(7) in (25).
We obtain

\begin{equation}
\phi(x) = \phi_{cl}(x) + \phi_{semi}(x)
\end{equation}

\noindent
where

\begin{equation}
\phi_{cl} = \frac{2}{\gamma} \int V'(x) dx  = \frac{2}{\gamma} V(x)
\end{equation}

\noindent
and

\begin{equation}
\phi_{semi}(x) = - \frac{2}{\gamma} 
\sum_{n\geq 1} \int \frac{x^{2n} (-1)^{3n+1}}{2^{2n}(2n)!} 
\frac{\partial^{2n+1} V}{\partial x^{2n+1}}  dx
\end{equation}

The existence of a potential for a {\it dissipative, semiclassical} dynamical system
is thus ascertained. The method essentially relies on a dynamical definition
$p_r$ as a derivative of the potential $\phi(x)$ in a system described by an
overdamped quantum Markov process in the weak noise and semiclassical limit. 
As elaborated 
earlier in Sec.III $p_r$ has a statistical origin which drives the
system away from its stable state $x_s$ to a preassigned remote state $x_f$
from which the system relaxes in absence of $p_r$. It is thus important
to realize that both the dynamical and statistical notions are kept
intact in the quantum treatment.

\section{Conclusions}

Keeping in view of the quantum nature of the system 
in terms of the Wigner's phase space function we examine the
semiclassical dynamics of a dissipative system in an Ohmic environment.
The weak noise limit
of the stochastic process then allows us to capture the essential
features of the dynamics within the framework of an auxiliary 
Hamiltonian description at the 
semiclassical level. Our results are summarized as follows :

\vspace{0.7cm}
\noindent
(i) The Wigner's quantum correction to classical Liouville equation gives rise
to an optimal force in addition
to usual the classical force term. This quantum optimal force is essentially
a result of an interplay of nonlinearity of the system and the thermal 
fluctuations of its environment and is derivable
in terms of an auxiliary Hamiltonian description.

\vspace{0.7cm}
\noindent
(ii) This term is also responsible for modification of growth and decay
of large fluctuations from equilibrium for the  appropriately
thermalized quantum system (compared to its classical counterpart).
The symmetry of the fluctuational and the relaxational paths, signifying 
the detailed balance, however, as expected is kept preserved.

\vspace{0.7cm}
\noindent
(iii) The quantum correction term implies the existence
of a potential for the dissipative semiclassical system.

\vspace{0.5cm}
Since the fluctuational and the relaxational paths have been experimentally
demonstrated as a part of physical reality by analogue experiments \cite{nature}
in the realm of large fluctuations, we believe that the essential 
modification of the integrable, nonintegrable and the singular
topological features of the dynamics due to  
semiclassical correction might be
relevant in several contexts. We hope to address some of these issues in a
future communication.

\acknowledgments
B. C. Bag is indebted to the Council of Scientific and
Industrial Research for a fellowship.

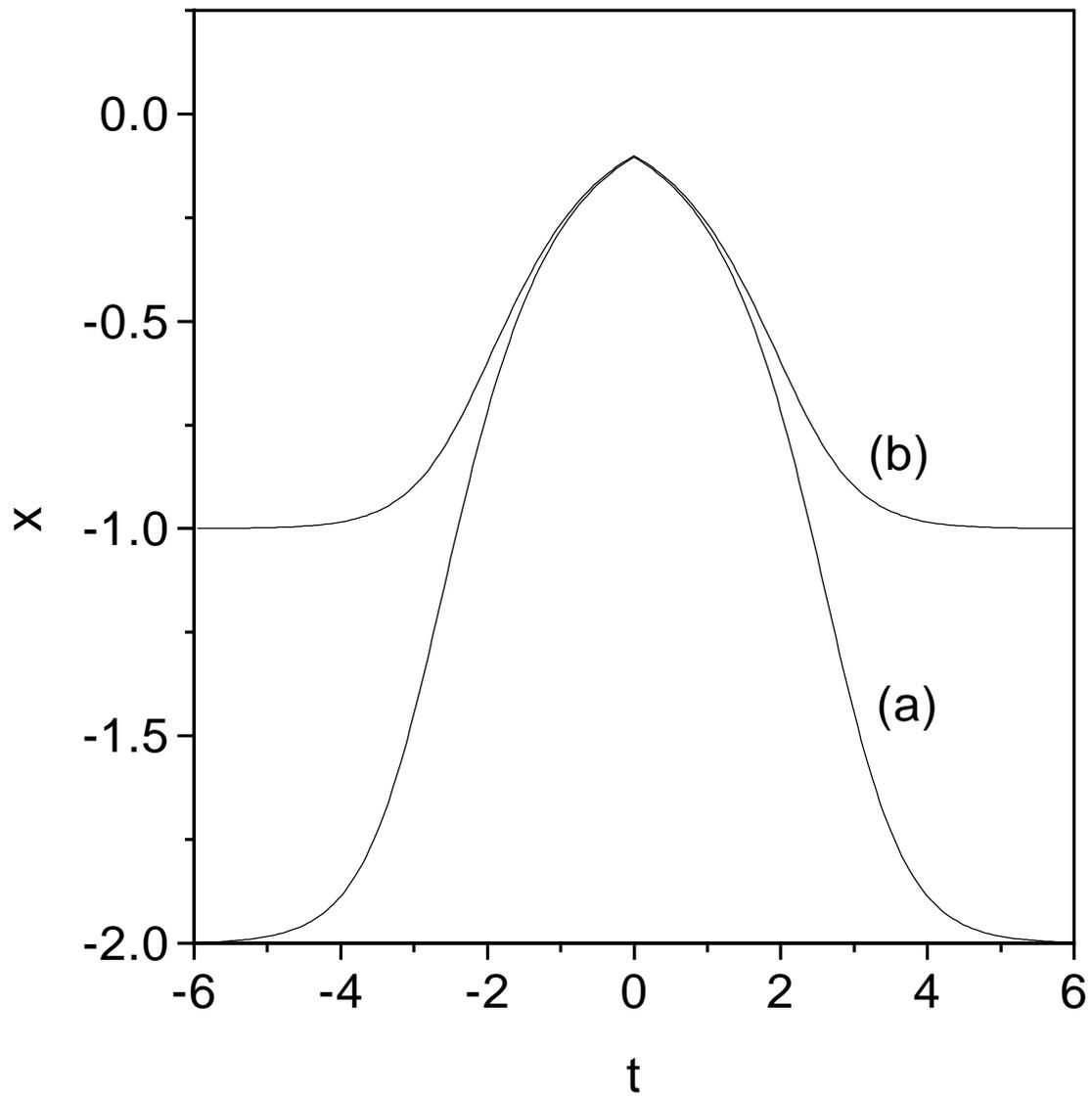
\begin{figure}
\caption{ A plot of system co-ordinate $x$ vs time $t$ signifying the
fluctuational and relaxational paths according to Eq.(8) for the
model system with $V(x) = \frac{1}{4} x^4 - \frac{1}{2} x^2$ for $(a)$ 
semiclassical case ($\epsilon = 1$) and $(b)$ classical case ($\epsilon = 0$), 
(units arbitrary).}
\end{figure}

\end{document}